\documentclass{aa}  

\usepackage{graphicx}
\usepackage{txfonts}
\usepackage[]{hyperref}
\begin{document}

   \title{A hidden molecular outflow in the LIRG Zw~049.057}
   \subtitle{}

   \author{N.~Falstad\inst{1}
          \and
          S.~Aalto\inst{1}
          \and
          J.~G.~Mangum\inst{2}
          \and
          F.~Costagliola\inst{1}
          \and
          J.~S.~Gallagher\inst{3}
          \and
          E.~Gonz\'alez-Alfonso \inst{4}
          \and
          K.~Sakamoto\inst{5}
          \and
          S.~K{\"o}nig\inst{1}
          \and
          S.~Muller\inst{1}
          \and
          A.~S.~Evans\inst{2,6}
          \and
          G.~C.~Privon\inst{7,8}
   }
          
  \institute{Department of Space, Earth and Environment, Chalmers University of Technology, Onsala Space Observatory,
     439 92 Onsala, Sweden \\
     \email{niklas.falstad@chalmers.se}
     \and
     National Radio Astronomy Observatory, 520 Edgemont Road, Charlottesville, VA 22903, USA
     \and
     Department of Astronomy, University of Wisconsin-Madison, 5534 Sterling, 475 North Charter Street, Madison WI 53706, USA
     \and
     Universidad de Alcal\'a de Henares, Departamento de F{\'i}sica, Campus Universitario, E-28871 Alcal\'a de Henares, Madrid, Spain
     \and
     Institute of Astronomy and Astrophysics, Academia Sinica, PO Box 23-141, 10617 Taipei, Taiwan
     \and
     Department of Astronomy, 530 McCormick Road, University of Virginia, Charlottesville, VA 22904, USA
     \and
     Instituto de Astrof{\'i}sica, Facultad de F{\'i}sica, Pontificia universidad Cat{\'o}lica de Chile, Casilla 306, Santiago, Chile
     \and
     Departamento de Astronom{\'i}a, Universidad de Concepci{\'o}n, Casilla 160-C, Concepci{\'o}n, Chile
  }

   \date{}

 
  \abstract
      { Feedback in the form of mass outflows driven by star formation or active galactic nuclei is a key component of galaxy evolution. The luminous infrared galaxy Zw~049.057 harbours a compact obscured nucleus with a possible far-infrared signature of outflowing molecular gas. Due to the high optical depths at far-infrared wavelengths, however, the interpretation of the outflow signature is uncertain. At millimeter and radio wavelengths, the radiation is better able to penetrate the large columns of gas and dust responsible for the obscuration.}
   {We aim to investigate the molecular gas distribution and kinematics in the nucleus of Zw~049.057 in order to confirm and locate the molecular outflow, with the ultimate goal to understand how the nuclear activity affects the host galaxy.}
   {We used high angular resolution observations from the Submillimeter Array (SMA), the Atacama Large Millimeter/submillimeter Array (ALMA), and the \emph{Karl G. Jansky} Very Large Array (VLA) to image the CO $J=2-1$ and $J=6-5$ emission, the $690$~GHz continuum, the radio centimeter continuum, and absorptions by rotationally excited OH.}
   {The CO line profiles exhibit wings extending ${\sim}300$~km\,s$^{-1}$ beyond the systemic velocity. At centimeter wavelengths, we find a compact (${\sim}40$~pc) continuum component in the nucleus, with weaker emission extending several $100$~pc approximately along the major and minor axes of the galaxy. In the OH absorption lines toward the compact continuum, wings extending to a similar velocity as for the CO are only seen on the blue side of the profile. The weak centimeter continuum emission along the minor axis is aligned with a highly collimated, jet-like dust feature previously seen in near-infrared images of the galaxy. Comparison of the apparent optical depths in the OH lines indicate that the excitation conditions in Zw~049.057 differ from those within other OH megamaser galaxies.}
   {We interpret the wings in the spectral lines as signatures of a nuclear molecular outflow. A relation between this outflow and the minor axis radio feature is possible, although further studies are required to investigate this possible association and understand the connection between the outflow and the nuclear activity. Finally, we suggest that the differing OH excitation conditions are further evidence that Zw~049.057 is in a transition phase between megamaser and kilomaser activity.}
   \maketitle
%

   \section{Introduction}
   Evidence is accumulating that a subset of luminous and ultraluminous infrared galaxies \citep[(U)LIRGS, see][for a review]{san96} harbor highly obscured dusty nuclei with optically thick continuum up to millimeter (mm) wavelengths \citep[e.g.][]{sak08,sak13,gon12,cos13,aal15,fal15,fal17,mar16}. These objects seem to be in a very active phase of evolution, their power supplied by central concentrations of molecular gas accreting onto a supermassive black hole (SMBH) or feeding intense star formation in the nuclear regions \citep[e.g.][]{gon12,cos13,aal16}. Some of these compact obscured nuclei (CONs) are also able to drive powerful molecular outflows at a rate which cannot be sustained for more than a few Myr \citep{aal12,cic14}. Yet, our understanding of how processes in the nucleus affect galaxy evolution is still limited; it is thus important both to determine the nature of the nuclear power source and to study the feedback processes that may eventually clear the obscuring material and reveal the activity in the nuclei of these galaxies.

    Due to the obscured nature of these objects, however, direct examination of their nuclear regions is impossible or impractical at many wavelengths. Since the dust opacity decreases with increasing wavelengths, the obscuring material is more easily penetrated with observations at radio or mm/sub-mm wavelengths where interferometry also offers a tool to achieve high angular resolution. Thus, although optical depth effects can still be an issue, imaging the continuum and molecular lines at these wavelengths can be used to probe the morphology, physical conditions, dynamics, and chemistry in obscured nuclei \citep[e.g.][]{sak08,aal09,var14,sco17,baa17}. With more information available about the nuclear environment it is possible to investigate the processes responsible for the high energy output and how they, over time, might affect the structure of the host galaxy.

\begin{figure}  
  \centering
  \includegraphics[width=8.0cm]{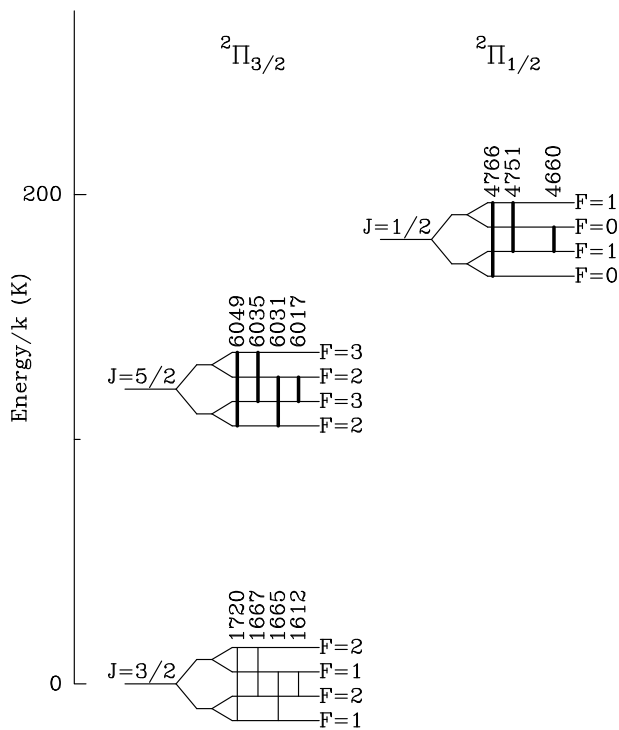}
  \caption{Energy diagram showing the ground and two first excited rotational states of OH. The $\Lambda$-doubling and hyperfine levels are not to scale. Transistions between hyperfine levels are indicated with lines connecting the levels in question and their frequencies in MHz are shown above the lines. Transitions observed in this paper are indicated with boldface lines.}
  \label{fig:energy}
\end{figure}

Observations of the hydroxyl radical (OH) have been a powerful method to probe outflow activity both in ULIRGs, using the low-lying far-infrared lines \citep[e.g.][]{fis10,stu11,vei13,gon17}, and in OH-megamasers, using the $1667$~MHz maser line \citep{baa89}. The OH molecule has a non-zero electronic angular momentum in its ground state and the interaction between the rotation of its nuclei and the orbital motion of the unpaired electron causes $\Lambda$-doublet splitting of the rotational energy levels. The levels are further split due to hyperfine interaction between the spin of the hydrogen atom and the orbital motion and spin of the electron. Transitions between the $\Lambda$-doublet pairs in the three lowest rotational levels give rise to spectral lines at frequencies close to $1.7$, $6.0$, and $4.7$~GHz where the obscuring properties of the dust are negligible. A diagram of these rotational levels with the $\Lambda$-doublet transitions indicated is shown in Fig. \ref{fig:energy}. In OH megamaser galaxies, the ground state lines, particularly the main lines at $1665$ and $1667$~MHz, are able to emit maser emission with luminosities more than six orders of magnitude higher than those of the most luminous galactic OH masers \citep[e.g.][]{baa82,hen90}. 
    
      \subsection{A highly obscured nucleus in Zw~049.057}
      Zw~049.057 is a moderately strong LIRG with an infrared luminosity of $L_{\mathrm{IR}}\approx1.8\times10^{11}$~$L_{\sun}$ \citep{san03}, at a distance of $56$~Mpc (linear scale: ${\sim}270$~pc/arcsec). It was described by \citet{sco00} as a galaxy with a highly inclined disk. We adopt a redshift of $z=0.012999$, corresponding to a velocity of $3897$~km\,s$^{-1}$, determined in the ESO nearby Abell cluster survey \citep{kat98} from optical absorption and emission lines. Throughout this paper the optical velocity convention is used with respect to the barycentric reference frame. All coordinates are given in the J2000.0 system. The CO $J=1-0$ emission of Zw~049.057 has been mapped at $2.7\arcsec$ ($730$~pc) resolution by \citet{pla91} using the Owens Valley millimeter-wave interferometer, revealing a compact distribution with most of the molecular gas located within ${\sim}0.4$~kpc (${\sim} 1.5\arcsec$) of the nucleus. A dust feature extending ${\sim}500$~pc (${\sim} 2\arcsec$) from the nucleus, along the minor axis, is seen in its NICMOS image \citep{sco00}.

      Zw~049.057 is known to host an OH megamaser in the low end of the megamaser luminosity range \citep{baa87,mar88}, but the exact nature of the nuclear power source has not been determined. Interestingly, in the sample of $77$ OH megamasers observed by \citet{mcb13}, it is one of only five sources with detected OH satellite lines at $1612$ and $1720$~MHz, and the only galaxy in which these lines are partially conjugate (when one line is seen in emission, the other is seen in absorption). As the conditions required to produce conjugate satellite lines differ from those that produce inversion in the main lines, \citet{mcb13} conclude that two separate inversion mechanisms exist in Zw~049.057. Furthermore, they suggest that the galaxy is in a transition phase between the OH megamasers and galaxies with lower luminosity masers. The galaxy nucleus has first been classified as starburst based on optical spectroscopy \citep{baa98}, but has then been reclassified as an AGN based on its radio activity \citep{baa06}. The main reason for this reclassification was the low spectral index ($\alpha=0.35$) between $1.4$ and $4.8$~GHz, while a classification purely based on the brightness temperature ($T_{\mathrm{B}}(4.8))=2.45\times10^{4}$~K) or q-factor ($q(4.8)=2.93$) would put it in the class of starburst nuclei. 
 
Recently, in their \emph{Herschel} far-infrared and submillimeter observations of Zw~049.057, \citet{fal15} found evidence of a highly obscured nucleus as well as a spectral signature of inflowing gas in the [\ion{O}{I}] line at $63$~$\mu$m. Although no outflow signatures were seen in the OH absorption lines, they did detect a tentative outflow signature in one of the H$_{2}$O lines. Moreover, in their IRAM Plateau de Bure observations, \citet{aal15} discovered possible continuum and self-absorption in the HCN and HCO$^{+}$ $J=3-2$ transitions, raising doubts as to their usefulness as probes of the nuclear environments in highly dust-obscured objects such as Zw~049.057.

With the aim of probing the kinematics and gas distribution in the nucleus, as well as to search for evidence of a hidden outflow not detected by {\emph Herschel}, we have used the \emph{Karl G. Jansky} Very Large Array (VLA) of the National Radio Astronomy Observatory (NRAO)\footnote{The National Radio Astronomy Observatory is a facility of the National Science Foundation operated under cooperative agreement by Associated Universities, Inc} to observe the $\Lambda$-doublet transitions in the first two rotationally excited states of OH, as well as the radio continuum at the frequencies of these transitions. We have also used the Submillimeter Array (SMA) \citep{ho04} on Mauna Kea, Hawaii, to observe the $J=2-1$ transition of CO. In addition we use archival CO $J=6-5$ data from the Atacama Large Millimeter/submillimeter Array (ALMA). In this paper we present and analyze the results of these VLA, SMA, and ALMA observations. The observation setups and data reduction processes are described in Sect. \ref{sec:observations}, the observational results are presented in Sect. \ref{sec:results}, and the implications of the results are discussed in Sect. \ref{sec:discussion}. Finally our conclusions are summarized in Sect. \ref{sec:conclusions}.
\section{Observations and data reduction}\label{sec:observations}

\subsection{VLA}
The observations were carried out in January 2017 (project 16B-432) in C-band ($4-8$~GHz) full polarization mode using the A-configuration of the array. There were 27 antennas in the array and the longest baseline was $36.4$~km. The total bandwidth was $2$~GHz, distributed over two basebands of $1$~GHz each, centered at $4.65$ and $6.0$~GHz. Each of these basebands was further divided into eight continuum subbands of $128$~MHz each. In addition, narrower subbands with higher spectral resolution were placed to observe the OH lines with rest frequencies $4660$, $4751$, $4766$, $6017$, $6031$, $6035$, and $6049$~MHz. The total on-source time for the observations was $2.4$~h and the final sensitivity achieved was $0.2$~mJy\,beam$^{-1}$ per $18$~km\,s$^{-1}$ (${\sim}250$~kHz) channel and $0.14$~mJy\,beam$^{-1}$ per $26$~km\,s$^{-1}$ (${\sim}500$~kHz) channel for the  $4.65$ and $6.0$~GHz data, respectively. No special astrometric calibration measurements were made and we estimate the astrometric accuracy to be $0.04\arcsec$ (${\sim}10$\% of the synthesized beam full width half maximum).

Data reduction was performed in the Common Astronomy Software Applications \citep[CASA;][]{mcm07} package using standard methods with 3C\,286 as bandpass and flux calibrator, and J1504+1029 as gain calibrator. Self-calibration of phase and amplitude was done using the continuum data, although this was not done when measuring the absolute position of the continuum peak (the continuum peak positions in the self-calibrated and non-self-calibrated images agree to within $0.01\arcsec$). The calibrated data was continuum subtracted in the \emph{uv}-plane using line-free channels and then imaged using Briggs weighting with robustness parameter $0.5$, resulting in a beam size of $0.44\arcsec \times 0.37\arcsec$ (PA ${\sim}-35\degr$) at $4.65$~GHz and $0.34\arcsec \times 0.28\arcsec$ (PA ${\sim}-40\degr$) at $6.0$~GHz. The resulting data cubes have a velocity resolution of $18-26$~km\,s$^{-1}$ and were cleaned to a threshold of three times the image rms noise in each velocity channel. A continuum image was created from all line-free channels using the multi-frequency synthesis algorithm with the same weighting as for the spectral line data cubes. This image was also cleaned to a threshold of three times the image rms noise. 

\subsection{SMA}
The SMA observations of Zw~049.057 were conducted on 2016 April 11 (project 2015B-S016). The array was in its extended configuration, with baselines ranging from $44$ to $226$~m, and the observations were carried out in $2$~GHz dual-receiver mode using the $230$ and $400$~GHz receivers. There were eight antennas in the array. The weather conditions were good with a $225$~GHz opacity of $\tau_{\mathrm{225}}=0.05$ and low wind throughout the observations. System temperatures were in the range $100-200$~K for the $230$~GHz receiver. We placed the redshifted CO $J=2-1$ line in the center of the upper sideband (USB) of the $230$~GHz receiver. The $230$~GHz continuum was obtained from the line-free channels of the USB of the $230$~GHz receiver. The total on-source time for the observations was $6.2$~h and the final sensitivity achieived was $10$~mJy\,beam$^{-1}$ per $15$~km\,s$^{-1}$ (${\sim}11.5$~MHz) channel.

Data reduction was performed in the CASA package after converting the SMA data into CASA measurement sets using the scripts \texttt{sma2casa.py} and \texttt{smaImportFix.py} provided by the SMA\footnote{The scripts are publicly available at www.cfa.harvard.edu/sma/casa.}. We employed standard methods to calibrate the data in CASA, using the quasar 3C\,273 as bandpass calibrator, the quasar J1504+104 as gain calibrator, and Ganymede as flux calibrator\footnote{The calibrator list for the SMA is available at http://sma1.sma.hawaii.edu/callist/callist.html.}. The astrometric accuracy was estimated by using the quasar J1549+026 as a test source, using the same calibration as for Zw~049.057, and turned out to be ${\sim}0.2\arcsec$. It is possible that our astrometric error is smaller than this because the quasars are $14\degr$ apart while the galaxy is only $4\degr$ away from J1504+104. The calibrated data was continuum subtracted in the \emph{uv}-plane using line-free channels and then imaged using Briggs weighting with robustness parameter $0.5$, resulting in a beam size of $1.34\arcsec \times 0.94\arcsec$ (PA ${\sim}-86\degr$). The resulting data cubes have a velocity resolution of $15$~km\,s$^{-1}$ and were cleaned to a threshold of three times the image rms noise in each velocity channel. A continuum image was created from line-free channels using the multi-frequency synthesis algorithm with the same weighting and was also cleaned to a threshold of three times the image rms noise. From the data cubes we generated maps of the velocity integrated intensity (moment 0), the intensity weighted velocity (moment 1),  and the intensity weigthed dispersion of the velocity (moment 2) of the spectral line. For the integrated intensity map, all channels of the spectral line were used without threshold. Masks were created by smoothing the data cube spatially and spectrally and then excluding any pixels with absolute values less than a threshold of $3 \sigma$. The masks were then applied to the unsmoothed data cubes before generating the higher moment maps. 

\subsection{ALMA} \label{subsec:obs_alma}
The CO $J = 6-5$ data set was obtained from the ALMA archive (project 2013.1.00524.S, PI N. Lu). The data were taken on 2015 May 19 with a total of 37 antennas in the $12$~m array, with baselines between $17.5$ and $555.5$~m. Four spectral windows each covered a bandwidth of $2$~GHz (${\sim}750$~km\,s$^{-1}$ at the frequency of CO $J=6-5$), with a frequency resolution of $15.6$~MHz. For calibration purposes data for the following sources were obtained during the course of the observations: \object{J1256-0547} as bandpass calibrator, \object{Titan} as flux calibrator, and \object{J1550+0527} as phase calibrator. The total on-source time for the observations was $272$~s and the final sensitivity achieved was $15$~mJy\,beam$^{-1}$ per $20$~km\,s$^{-1}$ (${\sim}46$~MHz) channel.

The CASA package was used for calibration and imaging of the data set. Since we were only interested in the position and the shape of the CO $J = 6-5$ emission, we made no major effort to improve the flux calibration with respect to the results from the pipeline. Thus, the flux calibration is uncertain to a factor of about 2. For the imaging we used briggs weighting with a robustness factor of 0.5. The resulting beam size is $0.23\arcsec\,\times\,0.21\arcsec$ (PA ${\sim}15\degr$). The astrometric accuracy was estimated by comparing the peak positions of the calibrators to their catalog positions and turned out to be $<0.05\arcsec$.

\section{Results}\label{sec:results}
\subsection{Radio cm continuum}\label{sec:continuum}
\begin{figure}  
  \centering
  \includegraphics[width=8.0cm]{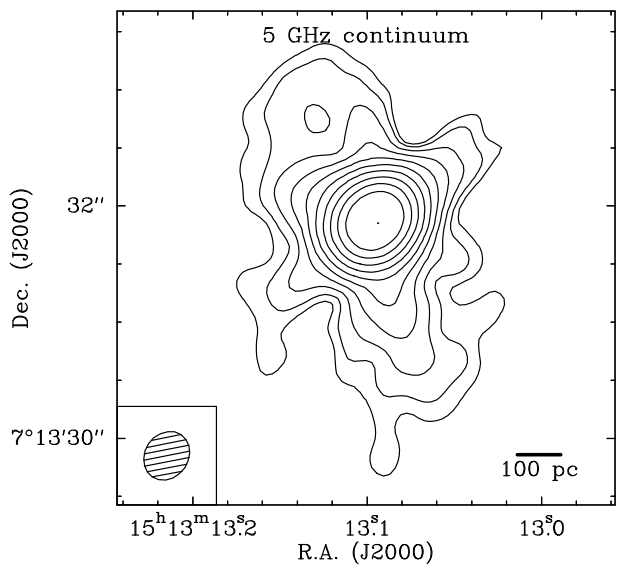}
  \caption{Radio continuum ($5.3$~GHz) contour map of Zw~049.057, produced from a combination of the $4.65$ and $6.0$~GHz data. Contours are at $1,2,4,8,16,32,64,128,256,512\times15$~$\mu$Jy\,beam$^{-1}$. The peak surface brightness is $25$~mJy\,beam$^{-1}$ and the off source rms is $5$~$\mu$Jy\,beam$^{-1}$. The fitted beam ($0.38\arcsec\times0.32\arcsec$, PA ${\sim -40\degr}$) is shown in the lower left corner.}
  \label{fig:VLAcontinuum}
\end{figure}

The radio continuum image (Fig. \ref{fig:VLAcontinuum}), produced from a combination of the $4.65$ and $6.0$~GHz data, is dominated by a bright compact component ($\mathrm{R.A.}=15^{\mathrm{h}}13^{\mathrm{m}}13^{\mathrm{s}}.095$, $\mathrm{decl.}=+07\degr13\arcmin31\arcsec.87$) in the middle of a more extended cross-shaped structure with an integrated flux which is $6-10\%$ of that in the core component. The extended structure consists of an elongated $900$~pc ($3.3\arcsec$, PA $\sim25\degr$) component along the major axis of the galaxy as well as a $700$~pc ($2.6\arcsec$, PA $\sim130\degr$) long structure approximately along the minor axis. The measured properties of the core as well as the total emission are summarized in Table \ref{tab:contflux}. The total flux density obtained for the continuum emission is consistent with the previous VLA measurements by \citet{baa06} and \citet{par10}.

\begin{table*} 
  \caption{Properties of the radio continuum emission.}             
  \label{tab:contflux}      
  \centering                          
  \begin{tabular}{l c c c c c c}        
  \hline\hline                 
 $\nu$ & $T_{\mathrm{B, peak}}$\tablefootmark{a} & Peak Brightness\tablefootmark{a} & $S_{\mathrm{core}}$\tablefootmark{a,b} & $S_{\mathrm{tot}}$\tablefootmark{c} & $\theta_{\mathrm{M}}\times\theta_{\mathrm{m}}$\tablefootmark{a,d} & PA\tablefootmark{a,e} \\ 
 (GHz) & ($10^{3}$~K) & (mJy\,beam$^{-1}$) & (mJy) & (mJy) & (mas) & ($\degr$)\\
  \hline                        
  $4.65$ & $9.4\pm0.3$ & $25.9 \pm 0.8$ & $30.8 \pm 0.9$ & $32.8 \pm 1.0$ & $183\times163\pm1$ & $119\pm3$ \\ 
  $6.00$ & $8.7\pm0.3$ & $24.3 \pm 0.8$ & $30.0 \pm 0.9$ & $32.6 \pm 1.0$  & $160\times141\pm1$ & $121\pm2$ \\ 
  $5.32\tablefootmark{f}$ & $8.9\pm0.3$ & $25.0 \pm 0.8$ & $30.4 \pm 0.9$ & $32.7 \pm 1.0$ & $171\times151\pm1$ & $120\pm3$ \\ 
    \hline                                   
  \end{tabular}
  \tablefoot{
    \tablefoottext{a}{Calculated from two-dimensional Gaussian fit to the continuum map using the CASA task IMFIT which deconvolves the clean beam from the fitted component size, see \citet{con97} for a discussion of the error estimates in this process. Uncertainties for the peak brightness and integrated flux density include an estimated $3$\%  calibration error added in quadrature to the noise error. The uncertainty for the deconvolved size does not include possible calibration errors.}
    \tablefoottext{b}{Integrated flux density of the core.}
    \tablefoottext{c}{Total integrated flux density of the core and the extended emission. The noise error was calculated by multiplying the image rms noise by the square root of the number of beam areas across the source.}
    \tablefoottext{d}{Core component deconvolved major $\times$ minor axis.}
    \tablefoottext{e}{Core component position angle.}
    \tablefoottext{f}{Produced from a combination of the $4.65$ and $6.0$~GHz observations.}
 } 
\end{table*}

\subsection{CO $J=2-1$ line emission}\label{sec:co}
\begin{figure}  
  \centering
  \includegraphics[width=8.0cm]{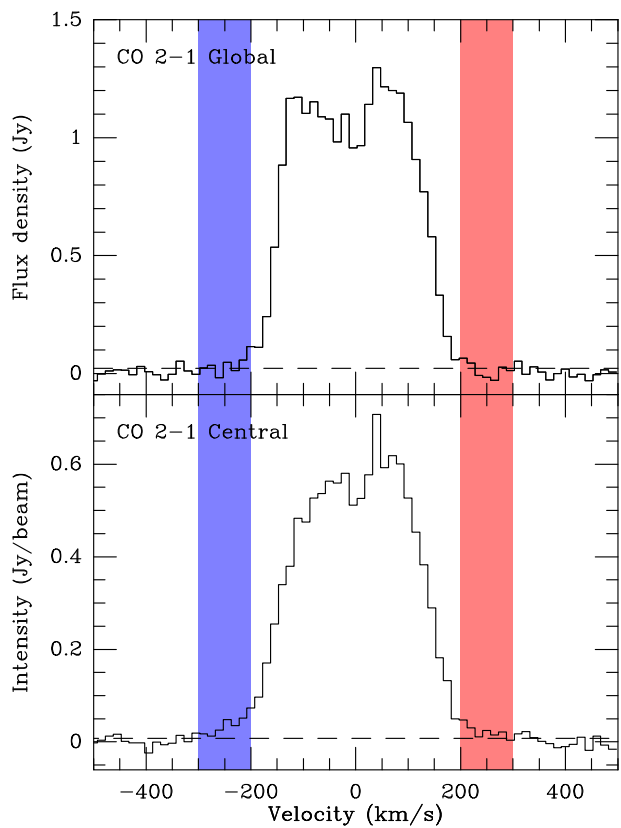}
  \caption{Spectra of CO $J=2-1$ in Zw~049.057. {\it Top:} integrated over the region with emission stronger than $3 \sigma$ in the integrated intensity map. {\it Bottom:} extracted from the peak of the integrated intensity map. The shaded regions indicate the velocity intervals of the blue- and red-shifted line wings seen in the lower spectrum. A dashed line in each spectrum indicates the $1 \sigma$ level of the noise above zero intensity. The spectra are referenced to the barycentric reference frame using $z=0.012999$.}
  \label{fig:co_spectra}
\end{figure}

Figure \ref{fig:co_spectra} shows two spectra of the CO $J=2-1$ transition, both the global spectrum integrated over the emitting region and the spectrum extracted at the peak of the integrated intensity map at $\mathrm{R.A.}=15^{\mathrm{h}}13^{\mathrm{m}}13^{\mathrm{s}}.094$, $\mathrm{decl.}=+07\degr13\arcmin31\arcsec.95$, which is slightly offset from the $230$~GHz continuum peak at $\mathrm{R.A.}=15^{\mathrm{h}}13^{\mathrm{m}}13^{\mathrm{s}}.096$, $\mathrm{decl.}=+07\degr13\arcmin31\arcsec.80$. The line flux computed by direct integration of the global spectrum is presented in Table \ref{tab:lineflux}. Compared to the single-dish observations reported by \citet{pap12} we recover approximately $55$\% of the CO $J=2-1$ line flux. The line profile in the central spectrum has weak wings on its blue- and red-shifted sides. The wings seem to originate in an unresolved region in the center of the nucleus and are not seen above the noise level in the global spectrum. 

\begin{figure*}  
  \centering
  \includegraphics[width=6.0cm]{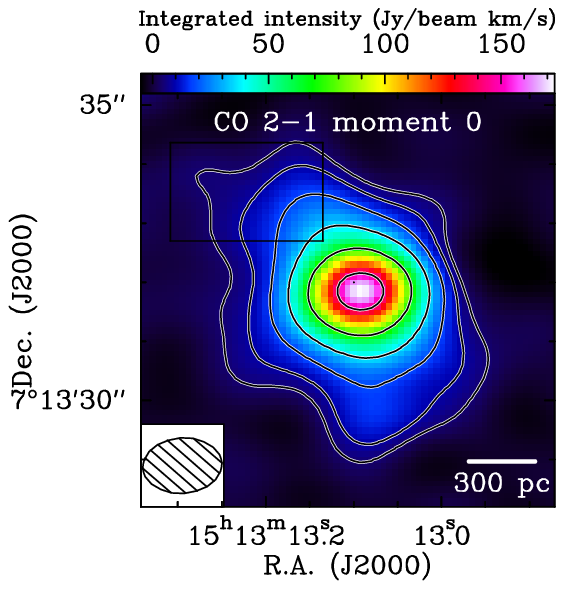}
  \includegraphics[width=6.0cm]{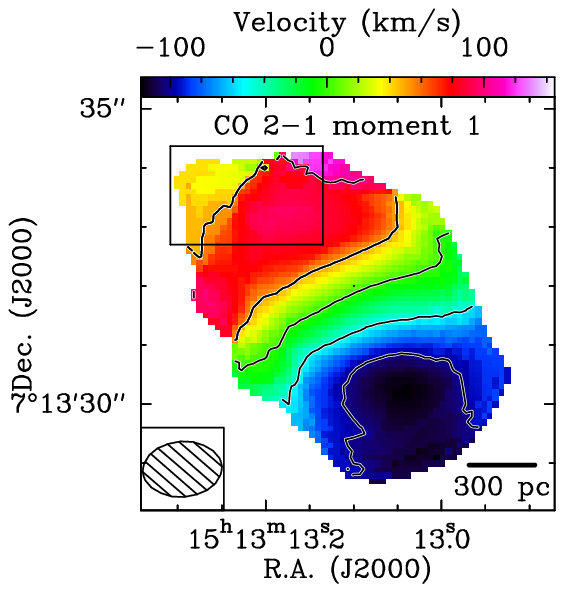}
  \includegraphics[width=6.0cm]{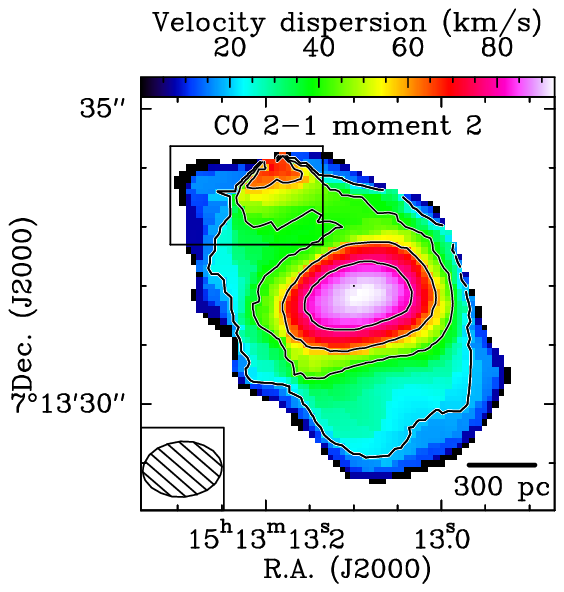}
  \caption{Moment maps of CO $J=2-1$. {\it Left:} integrated intensity map with contours at $1,2,4,8,16,32\times4.5$~Jy\,beam$^{-1}$\,km\,s$^{-1}$. {\it Center:} mean velocity map with contours at $-100, -50, 0, 50, 100$~km\,s$^{-1}$.  {\it Right:} dispersion map with contours at $20, 40, 60, 80$~km\,s$^{-1}$. The small box in each of the maps indicate the location of an extra emission component close to the systemic velocity. The fitted beam ($1.34\arcsec\times0.94\arcsec$, PA ${\sim -86\degr}$) is shown in the lower left corner of each panel.}
  \label{fig:co_moms}
\end{figure*}

\begin{figure*}  
  \centering
  \includegraphics[width=8cm]{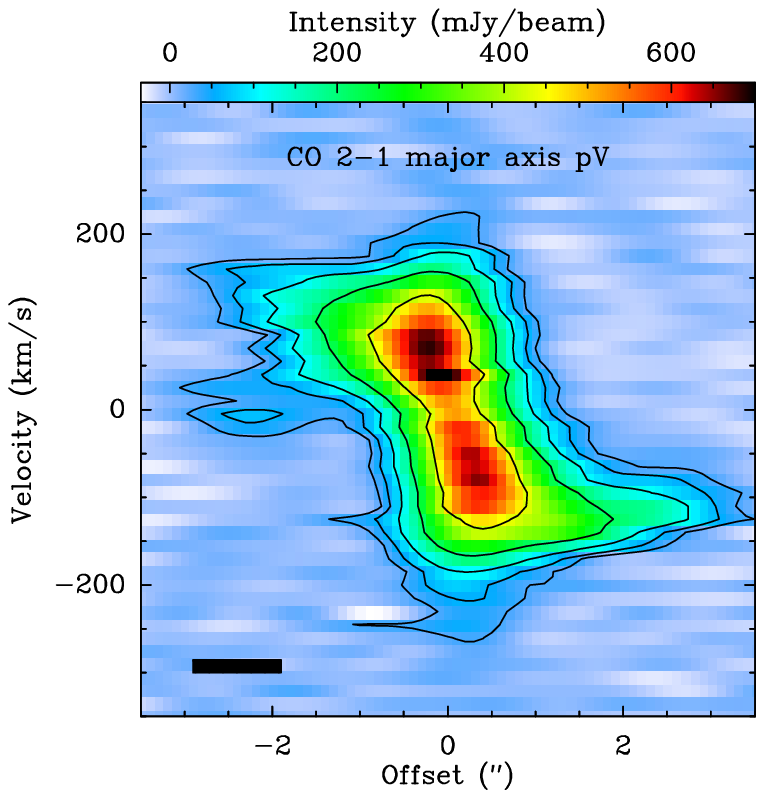}
  \includegraphics[width=8cm]{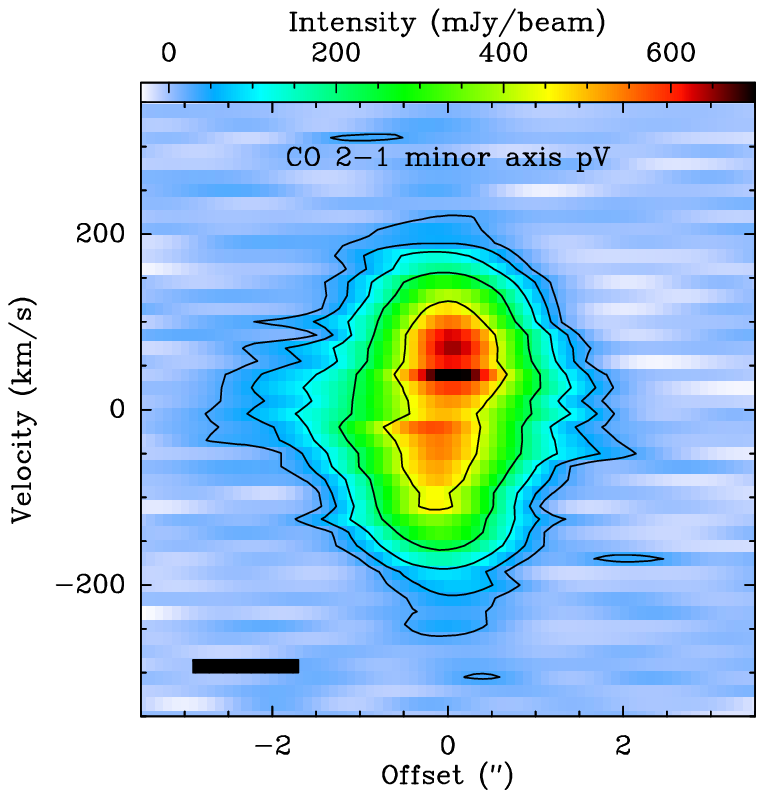}
  \caption{Position-velocity diagram for the CO $J=2-1$ line along the major (PA$\sim210\degr$, left panel) and minor (PA$\sim300\degr$, right panel) axis of the disk traced by the CO emission. The black rectangles in the bottom left corners show the spatial and velocity resolution of the data. Contours are at $1,2,4,8,16\times 27$~mJy\,beam$^{-1}$.}
  \label{fig:co_pv}
\end{figure*}

The integrated intensity, velocity field, and dispersion maps are presented in Fig. \ref{fig:co_moms} and the pV diagrams along the major and minor axes are shown in Fig. \ref{fig:co_pv}. Most of the emission emerges from a central region with radius ${\sim}150$~pc (${\sim} 0.6\arcsec$), but faint emission extends out to ${\sim}800$~pc (${\sim} 3\arcsec$) in a structure with position angle ${\sim}30\degr$. A Gaussian fit to the strong emission results in a central component with dimensions (full width half maximum, deconvolved from the beam) $350\times 250$~pc ($1.3\arcsec\times0.9\arcsec$, PA ${\sim}25\degr$). 
There is a clear velocity gradient along the major axis, with negative velocities to the southwest and positive to the northeast of the central region. The velocity dispersion is dominated by a peak of approximately $90$~km\,s$^{-1}$ in the central region, although this is at least partly due to beam smearing, with a typical dispersion of $20-60$~km\,s$^{-1}$ away from the center. An extra emission component close to the systemic velocity is present in the northeast, seen as a decrease in velocity and an increase in dispersion.

  \subsection{CO $J=6-5$ line and $690$~GHz continuum emission }\label{sec:co65}
\begin{figure}  
  \centering
  \includegraphics[width=8.0cm]{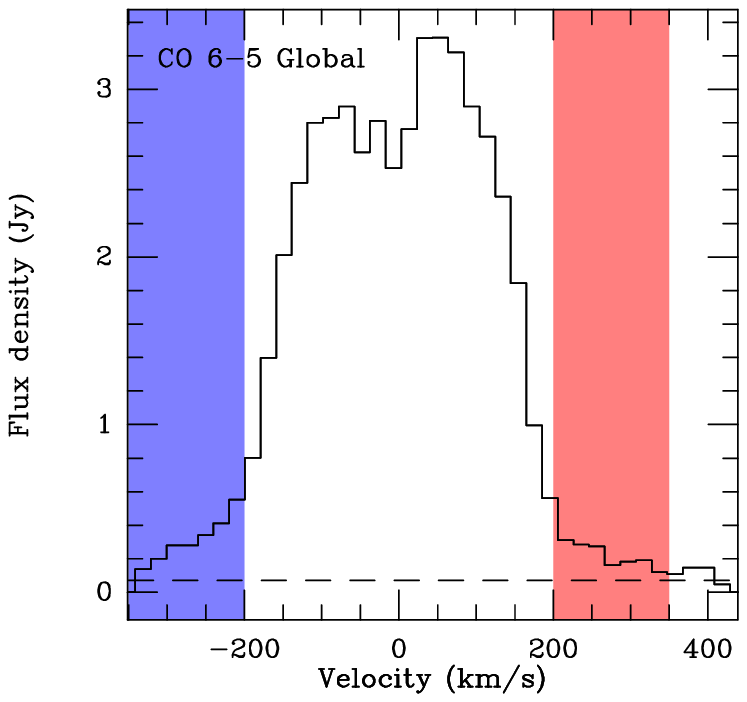}\\
  \includegraphics[width=8.0cm]{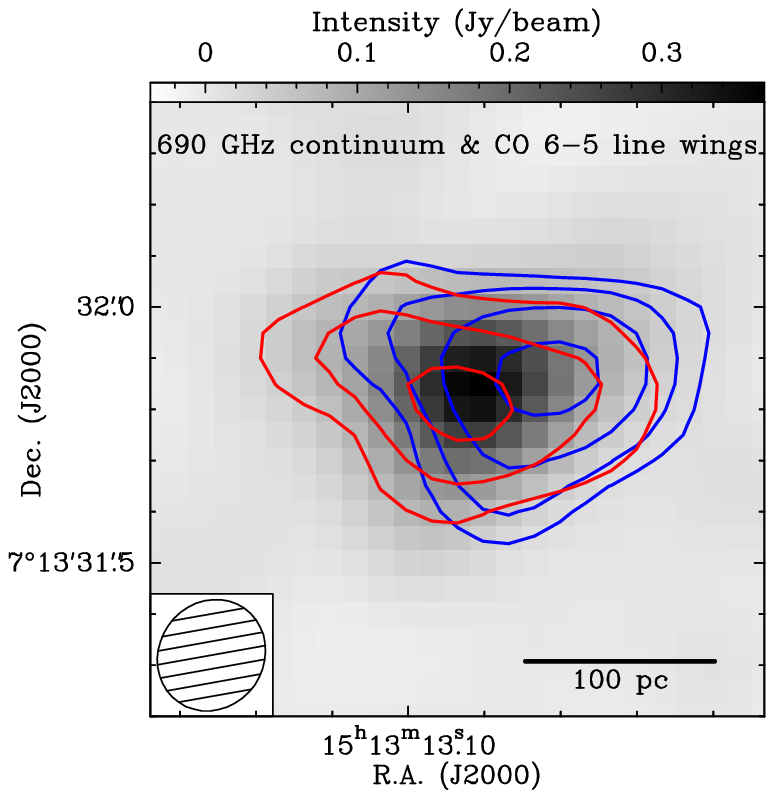}
  \caption{{\it Top:} Integrated spectrum of CO $J=6-5$ with the spectral regions used to make the figure in the bottom panel marked. A dashed line in the spectrum indicates the $1 \sigma$ level of the noise above zero intensity. The spectrum is referenced to the barycentric reference frame using $z=0.012999$. {\it Bottom:} Integrated intensity maps over the channels marked in the spectrum from the top panel overlaid on the $690$~GHz continuum greyscale image. Contours are given at $1, 2, 4, 8 \times 3\sigma$ for the wing emission. The fitted beam ($0.22\arcsec\times0.20\arcsec$, PA ${\sim -35\degr}$) is shown in the lower left corner.}
  \label{fig:outflow}
\end{figure}

  The top panel of Fig. \ref{fig:outflow} shows the CO $J=6-5$ spectrum integrated over the emitting region. There are clear line wings extending at least $350$~km\,s$^{-1}$ beyond the systemic velocity, on the blueshifted side the wing extends beyond the edge of the useable band. The bottom panel shows the integrated intensity maps taken over the line wings ($200-350$~km\,s$^{-1}$ beyond the systemic velocity), and the $690$~GHz continuum which peaks at $\mathrm{R.A.}=15^{\mathrm{h}}13^{\mathrm{m}}13^{\mathrm{s}}.092$, $\mathrm{decl.}=+07\degr13\arcmin31\arcsec.84$. The wing emission is oriented along an axis that is approximately perpendicular to the overall rotation of the CO $J=6-5$ and $J=2-1$ emission. 

\subsection{OH line absorption}\label{sec:oh}
\begin{figure*}
  \centering
  \includegraphics[width=18.0cm]{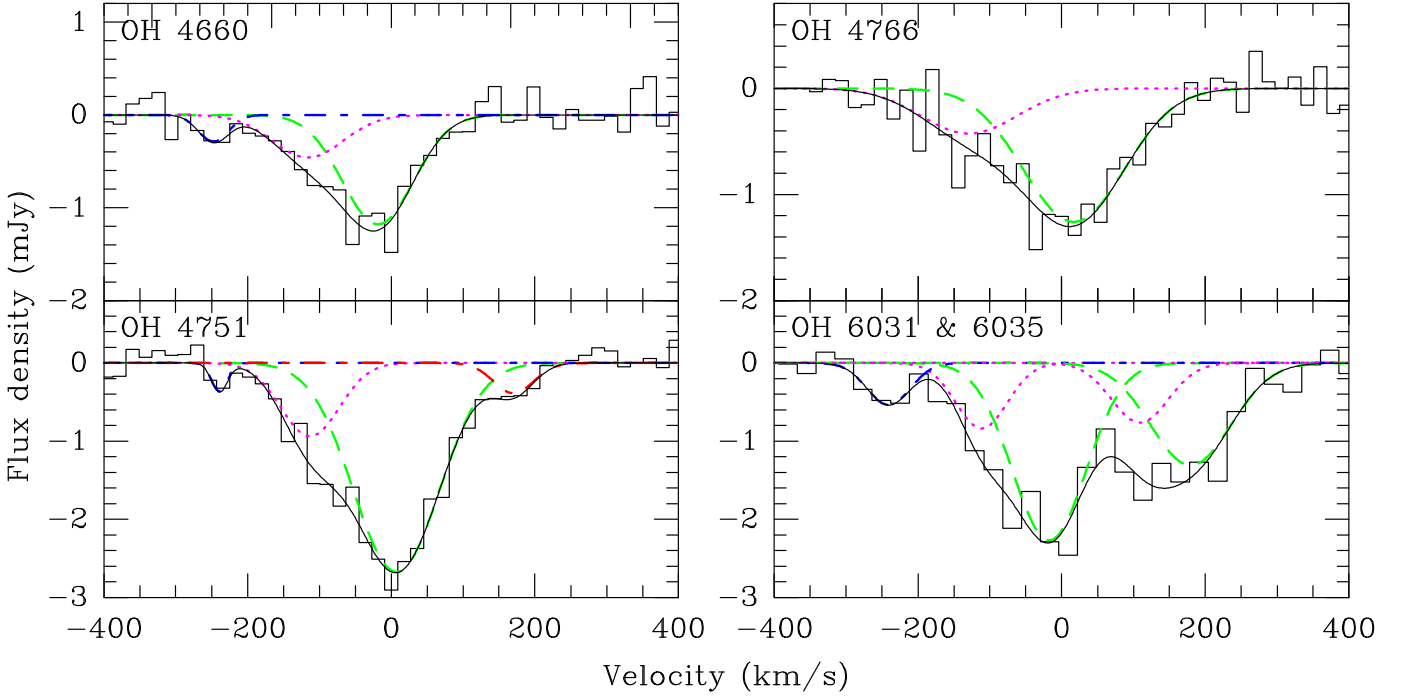}
  \caption{Detected OH absorption lines. The measured global spectra are plotted as histograms. The green dashed lines are Gaussian fits to the main feature in each spectrum. The magenta dotted lines are fits to the ${\sim}-110$~km\,s$^{-1}$ feature, except in the upper right spectrum where this component cannot be distinguished from the blueshifted line wing. The dashed-dotted blue (in all but the upper right panel) and red (in the lower left panel) lines are fits to the blue- and redshifted line wings, respectively. The black solid lines are the total fits. The velocity scale for the $6$~GHz lines is centered at the $6035$~MHz line and the rest velocity of the $6031$~MHz line is at $\sim 215$~km\,s$^{-1}$. All spectra are referenced to the barycentric reference frame using $z=0.012999$.}
  \label{fig:OHspectra2x2}
\end{figure*}

The global spectra of all detected lines are shown in Fig. \ref{fig:OHspectra2x2} together with a decomposition of the lines into multiple Gaussian components. The measured line fluxes obtained from direct integration of the line profiles are listed in Table \ref{tab:lineflux} together with the fitted parameters of the Gaussian components. The components that seem to be common to most lines are: a main component close to the systemic velocity, a weaker low velocity component around $-110$~km\,s$^{-1}$, and a blueshifted wing extending down to approximately $-300$~km\,s$^{-1}$. The different OH transitions are discussed in more detail in the following two sections. 

\begin{table*} 
  \caption{Line fluxes.}             
  \label{tab:lineflux}      
  \centering                          
  \begin{tabular}{l c c c c c c c}        
    \hline\hline                 
    Line &  $\nu_{\mathrm{rest}}$ & Component & $S_{\mathrm{line}}$\tablefootmark{a,b}  & $V_{\mathrm{cent}}$\tablefootmark{a}  & $V_{\mathrm {FWHM}}$\tablefootmark{a}   & $\int \! S_{\mathrm{line}}\,\mathrm{d}V$\tablefootmark{a,b} & $\tau_{\mathrm{app}}$\tablefootmark{c}  \\ 
    & (GHz) & & (mJy) & (km\,s$^{-1}$) & (km\,s$^{-1}$) & (mJy\,km\,s$^{-1}$) & \\
    \hline
    CO $J=2-1$ & $230.53800$ & Total & \ldots & \ldots & \ldots & $(341\pm51)\times10^{3}$ & \ldots \\
    \hline                        
    OH $4660$ & $4.66024$ & Total & \ldots & \ldots & \ldots & $-209\pm11$ & \ldots \\
    & & Main & $-1.18\pm0.34$ & $-18\pm12$ & $117\pm18$ & $-146\pm35$ & $0.039$ \\
    & & Low velocity &$-0.46\pm0.35$ & $-116\pm30$ & $116\pm54$ & $-57\pm35$ & $0.015$ \\
    & & Wing & $-0.29\pm0.38$ & $-247\pm14$ & $50\pm53$ & $-15\pm12$ & $0.009$ \\
    OH $4751$ & $4.75066$ & Total & \ldots & \ldots & \ldots & $-534\pm17$ & \ldots \\
    & & Main & $-2.67\pm0.18$ & $7\pm2$ & $142\pm7$ & $-404\pm18$ & $0.091$ \\
    & & Low velocity &$-0.94\pm0.10$ & $-113\pm4$ & $96\pm8$ & $-96\pm6$ & $0.031$ \\
    & & Wing & $-0.36\pm0.25$ & $-239\pm9$ & $26\pm13$ & $-10\pm5$ & $0.012$ \\
    & & Red wing & $-0.39\pm0.21$ & $170\pm11$ & $66\pm25$ & $-27\pm10$ & $0.013$ \\
    OH $4766$ & $4.76556$ & Total & \ldots & \ldots & \ldots & $-296\pm13$ & \ldots \\
    & & Main & $-1.26\pm0.11$ & $17\pm5$ & $166\pm11$ & $-222\pm12$ & $0.042$ \\
    & & Low velocity & $-0.43\pm0.17$ & $-129\pm18$ & $158\pm51$ & $-72\pm17$ & $0.014$ \\
    OH $6017$ & $6.01675$ & Total & $<0.19$\tablefootmark{d} & \ldots & \ldots & \ldots & \ldots \\
    OH $6031$ & $6.03075$ & Total &  \ldots & \ldots & \ldots & $-247\pm9$ & \ldots \\
    & & Main &$-1.31\pm0.11$ & $180\pm6$\tablefootmark{e} & $132\pm10$ & $-184\pm15$ & $0.045$ \\
    & & Low velocity & $-0.76\pm0.14$ & $109\pm7$\tablefootmark{e} & $94\pm14$ & $-76\pm13$ & $0.026$ \\
    OH $6035$ & $6.03509$ & Total & \ldots & \ldots & \ldots & $-410\pm14$ & \ldots \\
    & & Main & $-2.27\pm0.16$ & $-17\pm5$ & $121\pm7$ & $-292\pm11$ & $0.079$ \\
    & & Low velocity & $-0.84\pm0.33$ & $-111\pm3$ & $85\pm23$ & $-75\pm21$ & $0.028$ \\
    & & Wing & $-0.54\pm0.34$ & $-241\pm14$ & $71\pm35$ & $-41\pm16$ & $0.018$ \\
    OH $6049$ & $6.04909$ & Total & $<0.13$\tablefootmark{d} & \ldots & \ldots & \ldots & \ldots \\
    \hline                                   
  \end{tabular}
  \tablefoot{
    \tablefoottext{a}{Parameters for individual components are derived from Gaussian fits to the global spectra and total line fluxes are derived from direct integration of the line profile.}
    \tablefoottext{b}{The uncertainties in the VLA fluxes include an estimated $3$\% calibration error added in quadrature to the noise error. The calibration error in the SMA data is estimated to be $15$\%.}
    \tablefoottext{c}{Apparent optical depth $\tau_{\mathrm{app}}=-\ln(1+S_{\mathrm{line}}/S_{\mathrm{cont}})$. A covering factor of $100\%$ is assumed and $\tau_{\mathrm{app}}$ is thus a lower limit to the true optical depth.}
    \tablefoottext{d}{$1 \sigma$ upper limit.}
    \tablefoottext{d}{Relative to the $6035$~MHz line which is located ${\sim}215$~km\,s$^{-1}$ below the $6031$~MHz line.}
  }
\end{table*}

\subsubsection{4.7 GHz lines}
Both the main $4751$~MHz line and the two satellites at $4660$ and $4766$~MHz are detected in absorption toward the compact component of the continuum emission. The $4660$~MHz satellite line can be decomposed into a main component, a low-velocity component, and a wing component. Compared to the other lines, the profile of this line shows a lack of absorption on the red-shifted side, possibly due to maser emission filling in some of the absorption on the red side of the profile. The overall shape of the profile for the $4766$~MHz line seems similar to that of the $4660$~MHz line but as the blue side of its profile seems noisier it is hard to distinguish between a low-velocity and a wing component. The main $4751$~MHz line has the same basic components as the $4660$~MHz line, although the blue-shifted wing is relatively weaker. Unlike the other lines it also exhibits a red-shifted wing. The relative peak absorption depths of the lines are close to the ratios \citep[$1$:$2$:$1$ for the $4660$:$4751$:$4766$~MHz lines; e.g.][]{des77} expected under local thermodynamic equilibrium (LTE) while the relative line fluxes differ from these ratios due to the different profiles of the lines. This is possibly an effect of regions of differently excited OH gas not having the same continuum covering factor. 

\begin{figure*}  
  \centering
  \includegraphics[width=8cm]{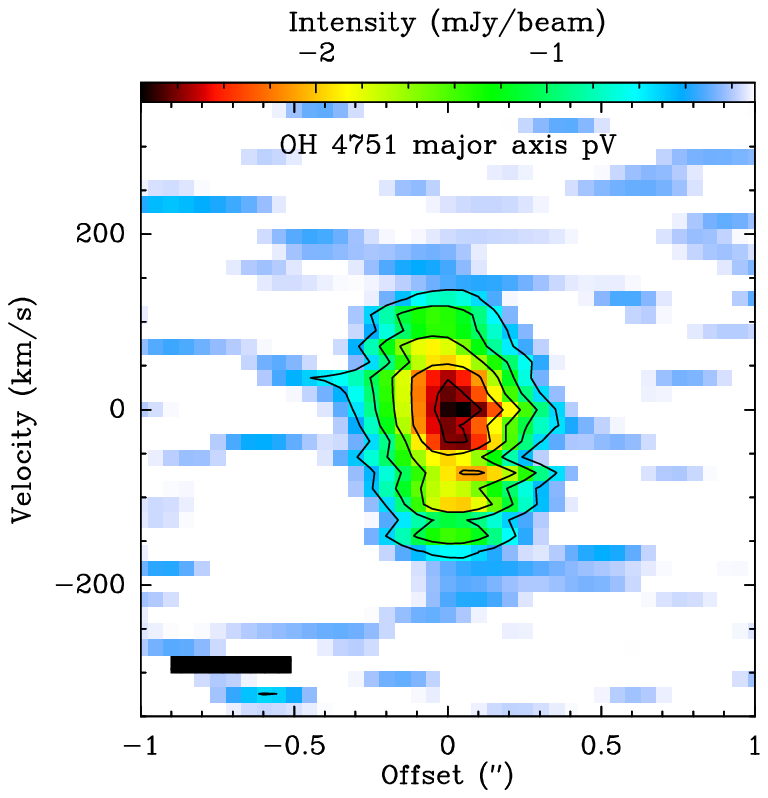}
  \includegraphics[width=8cm]{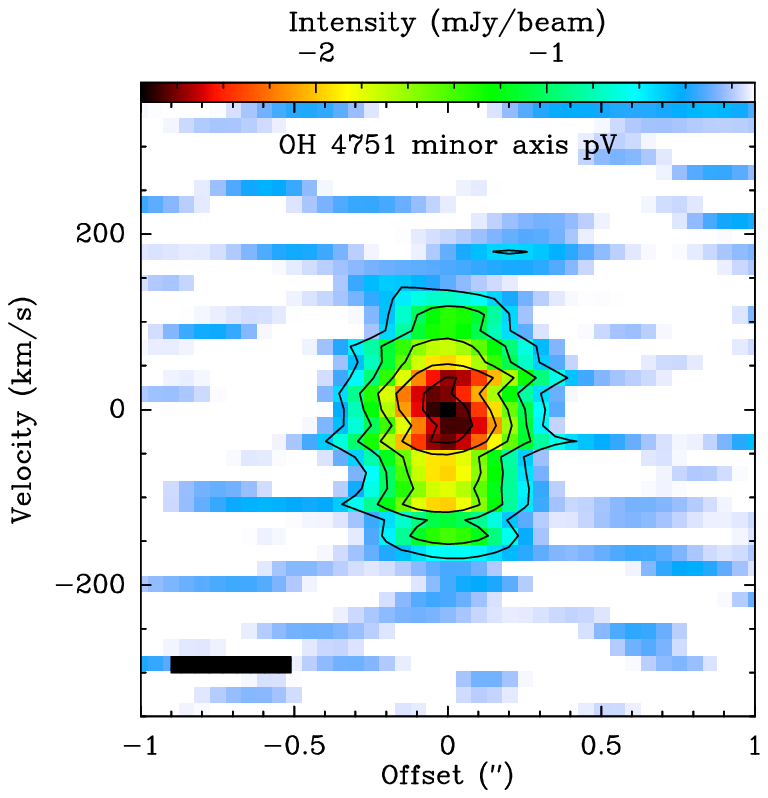}
  \caption{Position-velocity diagram for the OH $4751$~MHz line along the major (left) and minor (right) axis continuum features with position angles $\sim205\degr$ and $\sim310\degr$, respectively. The black rectangles in the bottom left corners show the spatial and velocity resolution of the data. Contour steps are $0.51$~mJy\,beam$^{-1}$ ($3 \sigma$).}
  \label{fig:4751pv}
\end{figure*}

\begin{figure}  
  \centering
  \includegraphics[width=8.0cm]{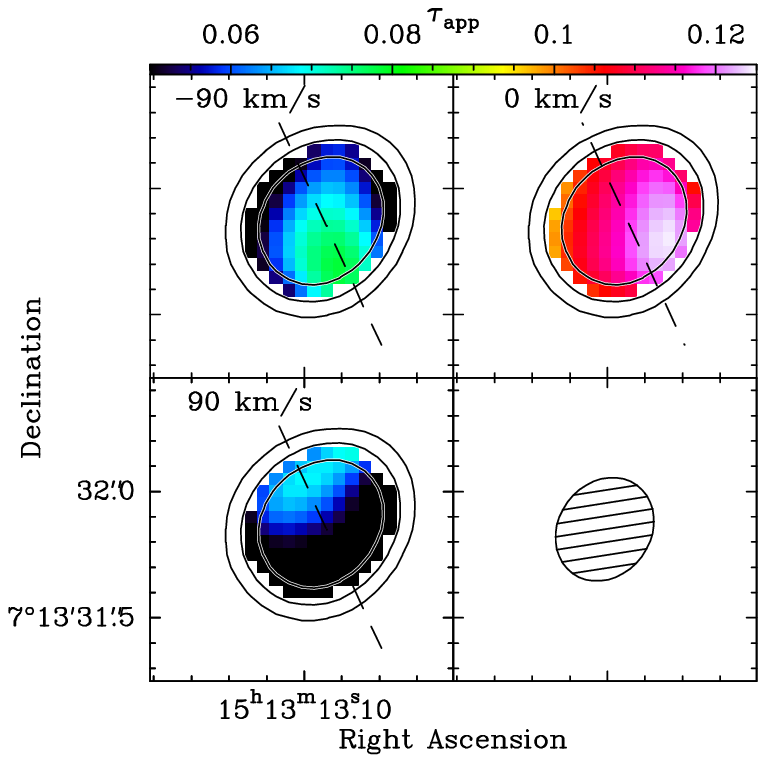}
  \caption{Channel maps of the apparent optical depth in the OH $4751$~MHz line. Note that channels have been averaged together in groups of five to produce bins of width $90$~km\,s$^{-1}$. The three highest contours from the continuum map in Fig. \ref{fig:VLAcontinuum} are included for reference. The dashed line in the first three panels indicates the major axis of the extended radio continuum structure. The lower right panel shows the fitted beam.}
  \label{fig:chanmap_4751}
\end{figure}

Position-velocity (pV) diagrams for the OH $4751$~MHz lines along the major and minor axis continuum structures are shown in Fig. \ref{fig:4751pv}. A velocity gradient along the major axis is seen in the pV-diagram (left panel). No velocity shift is seen along the minor axis (right panel) but a tentative, marginally resolved red-shifted (${\sim}200$~km\,s$^{-1}$) feature appears approximately $0.2\arcsec$ to the northwest of the continuum peak. After averaging together channels in groups of five to produce $90$~km\,s$^{-1}$ bins, maps of the apparent optical depth ($\tau_{\mathrm{app}}=-\ln(1+S_{\mathrm{line}}/S_{\mathrm{cont}})$) in the $4751$~MHz line were created. The maps of the three central $90$~km\,s$^{-1}$ channels are shown in  Fig. \ref{fig:chanmap_4751}.

\subsubsection{6 GHz lines}
\begin{figure*}  
  \centering
  \includegraphics[width=8.0cm]{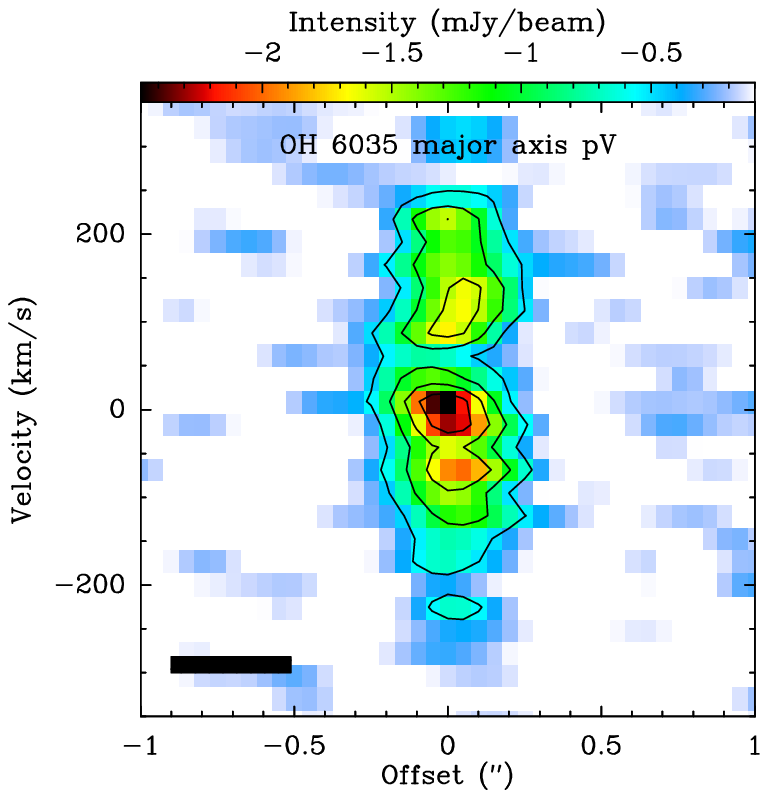}
  \includegraphics[width=8.0cm]{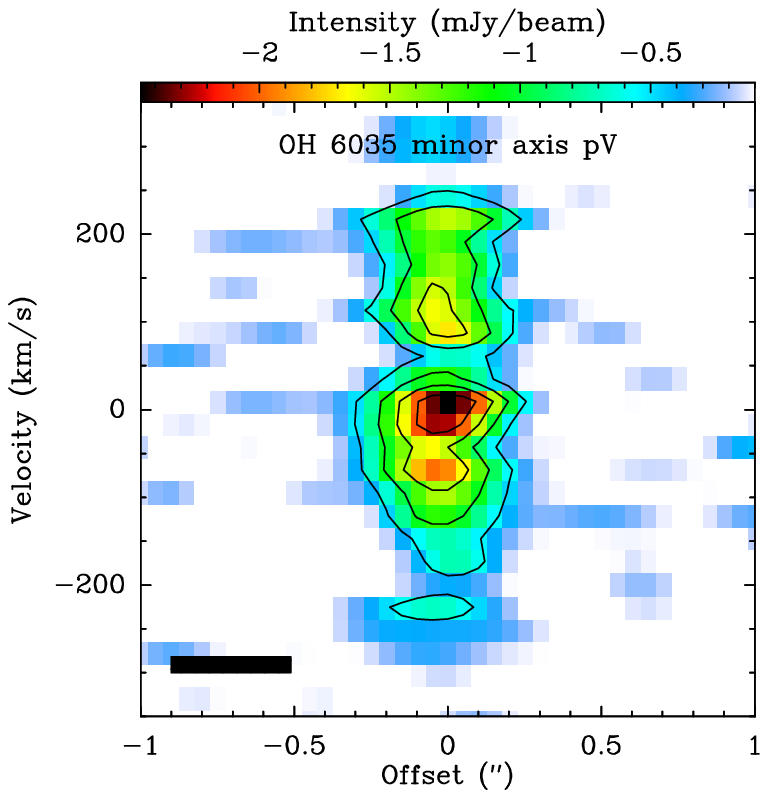}
  \caption{Position-velocity diagram for the OH $6031$ and $6035$~MHz lines along the major (left) and minor (right) axis continuum features with position angles $\sim205\degr$ and $\sim310\degr$, respectively. The velocity scale is centered at the $6035$~MHz line and the rest velocity of the $6031$~MHz line is at $\sim 215$~km\,s$^{-1}$. The black rectangles in the bottom left corners show the spatial and velocity resolution of the data. Contour steps are $0.39$~mJy\,beam$^{-1}$ ($3 \sigma$).}
  \label{fig:6000pv}
\end{figure*}

The two main lines at $6031$ and $6035$~MHz are detected in absorption but the satellites at $6017$ and $6049$~MHz are not. The $6035$~MHz line has a line wing similar to those in the satellite lines of the $4.7$~GHz triplet, while any line wing of the $6031$~MHz line would be blended with the main absorption component of the $6035$~MHz line. The relative peak absorption depths of the main lines and the upper limits to the satellite lines are consistent with the ratios expected under LTE conditions \citep[$1$:$14$:$20$:$1$ for the $6017$:$6031$:$6035$:$6049$~MHz lines; e.g.][]{des77}. The pV-diagrams, including both $6$~GHz lines, along the major and minor axis continuum structures are shown in Fig. \ref{fig:6000pv}. Both diagrams clearly show the blue wing below $-200$~km\,s$^{-1}$ in the $6035$~MHz line; along the major axis (left panel) it seems to be centered on the compact continuum component while it is possibly skewed toward the southeast (negative offset) when the slice is taken along the minor axis continuum feature (right panel).

\section{Discussion}\label{sec:discussion}

\subsection{Disk orientation and kinematics}\label{sec:dynamics}
To first order, under the assumption that the disk is circular and thin and that there are no non-circular motions, the inclination, $i$, of a disk is given by: $\cos{i}=b/a$, were $b/a$ is the ratio of the apparent minor and major axes. The CO moment maps (Fig. \ref{fig:co_moms}) indicate that the molecular gas is distributed in an inclined disk with projected dimensions $1.6\times1.2$~kpc ($5.9\arcsec\times4.4\arcsec$, PA ${\sim}30\degr$); this geometry is consistent with the near-infrared images of the nucleus presented by \citet{sco00}. The inclination as determined from the axis ratio of the CO disk is $\sim40\degr$.

The overall velocity gradient along the major axis (see Figs. \ref{fig:co_moms} and \ref{fig:co_pv}) gives the rotation, with the gas receeding in the northeast and approaching in the southwest with a maximum deprojected velocity of ${\sim}180$~km\,s$^{-1}$. Comparing with the rotation traced by the pV-diagrams of the strongest unblended OH absorption line (OH $4751$, see Fig. \ref{fig:4751pv}) it seems that the OH and CO probe regions with similar rotation, although on different scales. An extra, redshifted, component is seen in the $4751$~MHz absorption line (see Fig. \ref{fig:OHspectra2x2}). Based on the  pV-diagrams of Fig. \ref{fig:4751pv}, this extra absorption component seems to be located in the northwestern part of the nucleus as the only significant absorption above $150$~km\,s$^{-1}$ is seen in the pV-diagram taken along the minor axis continuum feature (right panel). This redshifted component is also discussed briefly in Sect. \ref{sec:outflow}.

The maps of the apparent optical depth at blueshifted, central, and redshifted velocities shown in Fig. \ref{fig:chanmap_4751} are also consistent with a rotation along the major axis. Below the systemic velocity (upper left panel of Fig. \ref{fig:chanmap_4751}), the optical depth peaks in the southwestern part of the nucleus while the peak at velocities above systemic (lower left panel of Fig. \ref{fig:chanmap_4751}) is located in the northeastern part. Interestingly, the peak optical depth at the systemic velocity (upper right panel of Fig. \ref{fig:chanmap_4751}) is not located in the center of the nucleus but is rather situated in the southwestern part of the nucleus. This could possibly be a signature of an inclined disk or ring-like structure where the far-side either is obscured in some way or has less background continuum to absorb; a situation similar to the one in III~Zw~35 (see \citet{pih01} and \citet{par05}). However, considering the spatial resolution of our observations, this remains rather tentative and would have to be tested with, for example, very long baseline interferometry (VLBI) observations of the maser lines at $1665$ and $1667$~MHz.

\subsection{Excitation conditions in the nucleus}\label{sec:physical}
It is intriguing that the main line at $4751$~MHz, which is the strongest one in the other absorption components, has a relatively weak blue wing compared to the other lines. A possible explanation for this behavior is that the main and blueshifted absorption components are formed in regions with different excitation conditions. This is what was seen in the megamaser galaxy Arp~220 where \citet{baa87} identified at least two main emission regions for the ground state maser lines: ``Region I'' which is a disk structure around the main radio component of the galaxy and is responsible for the absorption at systemic velocity, and ``region II'' which is associated with the eastern component of the nuclear source and is responsible for emission centered at a higher velocity. The same regions were also used by \citet{hen87} to explain why the $4660$~MHz line has a larger velocity and linewidth than the main $4751$~MHz line. Zw~049.057 has no known double structure in its radio continuum but it is possible that the blue wing originates in an outflow which creates a region II-like environment as it collides with the surrounding interstellar medium. 

\citet{baa87} detected the $1667$~MHz maser line in Zw~049.057 with a peak flux density of $12.5$~mJy on top of a $40$~mJy continuum, resulting in an apparent optical depth of ${\sim}0.27$. Assuming that the ground state maser line and the main absorption components in the excited lines originate in the same volume of gas in front of the ${\sim}40$~pc (${\sim}0.15\arcsec$) continuum source, and equating the absolute values of the excitation temperatures ($T_{\mathrm{ex}}$) between the hyperfine levels, it is possible to estimate the excitation temperatures of the rotational transitions, $T_{\mathrm{rot}}$, from the Boltzmann relation:
\begin{equation}\label{eq:boltzmann}
\frac{N_{\mathrm{upper}}}{N_{\mathrm{lower}}}=\frac{g_{\mathrm{upper}}}{g_{\mathrm{lower}}}\exp\left(\frac{-\Delta E}{kT_{\mathrm{rot}}}\right),
  \end{equation}
and the equations for the total populations of the states. The latter can be expressed as \citep[e.g.][]{hen86,hen87}:
\begin{equation}\label{eq:total_pop1}
  N_{\mathrm{18}}=2.4\times10^{14}\,T_{\mathrm{ex}}\,\Delta V_{\mathrm{FWHM}}\,\tau_{1667},
\end{equation}
 for the $^{2}\Pi_{3/2}\,J=\frac{3}{2}$ state,
\begin{equation}\label{eq:total_pop2}
  N_{\mathrm{5}}=1.6\times10^{14}\,T_{\mathrm{ex}}\, \Delta V_{\mathrm{FWHM}}\, \tau_{6035},\\
\end{equation}
for the $^{2}\Pi_{3/2}\,J=\frac{5}{2}$ state, and
\begin{equation}\label{eq:total_pop3}
  N_{\mathrm{6}}=1.6\times10^{14}\,T_{\mathrm{ex}}\, \Delta V_{\mathrm{FWHM}}\, \tau_{4751},
\end{equation}
for the $^{2}\Pi_{1/2}\,J=\frac{1}{2}$ state. Comparing the optical depths of the lines in the ground and rotationally excited states in Arp~220 in this way, \citet{hen86} and \citet{hen87} found excitation temperatures of $44$ and $74$~K for the transitions from the $^{2}\Pi_{3/2}\,J=\frac{5}{2}$ and $^{2}\Pi_{1/2}\,J=\frac{1}{2}$ states to the ground state. They note that these temperatures are close to the dust temperature of $61$~K in Arp~220 obtained by \citet{eme84} based on far-infrared data. For the transition between the $^{2}\Pi_{1/2}\,J=\frac{1}{2}$ and ground states, \citet{hen87} find similar values also in the other galaxies (NGC~3690, Mrk~231, and Mrk~273) of their sample.

Following the same analysis as \citet{hen86,hen87} in Zw~049.057, we obtain a rotational temperature of $60$~K for the $119$~$\mu$m (${\sim}2510$~GHz) transition from the $^{2}\Pi_{3/2}\,J=\frac{5}{2}$ state, close to the dust temperatures derived by \citet{fal15} based on a radiative transfer analysis of the far-IR rotational lines. However, for the $79$~$\mu$m (${\sim}3790$~GHz) transition from the $^{2}\Pi_{1/2}\,J=\frac{1}{2}$ state we obtain a considerably higher rotational temperature of $232$~K. The large difference compared to the temperatures found in other megamaser galaxies indicates either that one or more of the assumptions underlying the analysis are invalid or that the excitation conditions in Zw~049.057 are different. Indeed, in their study of rotationally excited OH in OH megamaser galaxies, \citet{hen87} found that the apparent optical depth of the $4751$~MHz line lay between $0.004$ and $0.04$, less than half the value that we find in Zw~049.057, in all of their sample galaxies. When comparing with the ground state lines they found a ratio between the $1667$ and $4751$~MHz lines of $-25\pm10$, approximately ten times as large as the ratio of $-2.7$ that we find. At the same time we find an apparent optical depth in the $6035$~MHz line of $0.079$, in line with the value of $0.1$ found in Arp~220 by \citet{hen86}. Assuming that the true rotational temperature between the ground and lowest rotational states should be close to the dust temperature, this indicates that the absolute excitation temperatures between the hyperfine levels in the $^{2}\Pi_{1/2}\,J=\frac{5}{2}$ and ground states might be similar to each other but different from those in the $^{2}\Pi_{3/2}\,J=\frac{1}{2}$ state.

The high apparent optical depth in the $4751$~MHz line clearly sets Zw~049.057 apart from other megamaser galaxies. An interesting comparison can instead be made with IC~860, another highly obscured galaxy \citep{aal15}. It is a partial maser in which the $1665$ and $1667$~MHz lines are dominated by absorption \citep{sch86} but with enough emission to classify it as a kilomaser \citep{baa89b,hen90}. Together with the weak megamasers Zw~049.057 and IRAS~11506-3851, IC~860 is located in the region of OH absorbers, rather than emitters, in the plot by \citet{baa89b} of infrared luminosity against the ratio of $25$ and $100$~$\mu$m flux densities (Fig. $2$ of \citet{baa89b}. In the measurements reported by \citet{man13}, the apparent optical depth of the OH $4751$~MHz line in IC~860 is even higher (${\sim}0.2$) than in Zw~049.057. This is possibly an indication that Zw~049.057 is in a transition phase between the megamasers and the kilomasers, with IC~860 on the weaker kilomaser side of this transition, and on its way to become a pure OH absorber in a process similar to the scenario suggested by \citet{hen90}. In this scenario, intensive star formation causes the strong infrared emission which is required for the megamaser activity. As the infrared luminosity decays, the galaxy then goes through a kilomaser stage before it finally ends up as an OH absorber.

\subsection{A nuclear outflow}\label{sec:outflow}
All detected OH absorption lines exhibit signatures of outflowing gas in the form of blueshifted wings in their line profiles toward the center of the nucleus. The maximum velocity extent of these wings is typically ${\sim}-300$~km\,s$^{-1}$ and they all, with the possible exception of the wing in the $6035$~MHz line, seem to originate in an unresolved region in the nucleus. Further evidence for the presence of a nuclear outflow comes from the line wings in the CO $J=2-1$ and $J=6-5$ emission lines toward the center of Zw~049.057 (see Figs. \ref{fig:co_spectra} and \ref{fig:outflow}). As the CO is seen in emission, no background continuum is required and wings are seen on both sides of the profile, extending out to velocities of at least $\pm300$~km\,s$^{-1}$ relative to the systemic velocity. Although this could also be interpreted as a signature of rapidly rotating gas in the center of the the nucleus, the outflow interpretation is supported by the blueshifted wings in OH which extend to approximately the same velocity as the blueshifted wing in the CO emission. The spatial orientation and shape of the CO $6-5$ line wings indicate a high-density narrow outflow. It might be related to the OH outflow, but further studies of the OH gas are required to confirm this. Interestingly, relative to the line core, the wings of the CO $J=6-5$ line are stronger than those of the CO $J=2-1$. This could be an indication that the excitation conditions are different in the outflow and the disk, which could in turn be used to uncover the mechanism driving the outflow. However, the large difference in angular resolution between the CO $J=6-5$ and $J=2-1$ observations complicates the matter, precluding a detailed analysis.

A tentative redshifted feature is also seen in the OH line at $4751$~MHz in the northwestern part of the nucleus. This feature is not seen in the nuclear spectrum of the line and not at all in the other lines. A possible reason for this is that the background continuum away from the absolute center of the galaxy is too weak for shallow wings to be detected over the noise, but differences in excitation conditions may also be important. The interpretation of this wing is less clear than for the blue wings in the center, and it is unclear if, and in that case how, it is related to the nuclear outflow. It is possible that it is a signature of gas flowing in toward the nucleus on a larger scale than the OH outflow, which is only seen toward the compact component of the radio continuum. Such an inflow would however be opposite to the kinematics traced by the CO (see Fig. \ref{fig:outflow}). Another possibility is that the feature comes from an unrelated process, for example an inner spiral arm passing in front of the radio continuum.

Strong molecular outflows in OH megamaser galaxies have been detected already by \citet{baa89} who found terminal velocities in the OH $1667$~MHz line as high as $800$~km\,s$^{-1}$ in the most luminous galaxies. The $1667$~MHz line in Zw~049.057 was found to have a terminal velocity of $260$~km\,s$^{-1}$, only slightly lower than the velocity we see in the rotationally excited lines. This is considerably lower than the outflow velocities detected in some other sources, but in line with the result that the outflow velocities are higher in more infrared-luminous galaxies \citep{baa89}.

\subsubsection{Hidden outflows in obscured nuclei}
The radio transitions of OH presented here can be compared to the far-infrared OH lines reported by \citet{fal15}. Interestingly, unlike the radio lines, the far-infrared lines do not show any obvious absorption extending down to projected velocities of $-300$~km\,s$^{-1}$. This is a similar situation to that observed in Arp~220, where an outflow with a projected terminal velocity of $800$~km\,s$^{-1}$ is seen in the $1667$~MHz OH line \citep{baa89}. This is much larger than the velocity implied by the outflow signature seen in the far-IR OH lines reported by \citep{gon12}, something that they interpret as evidence for dust extinction of the high-velocity gas combined with deceleration of the outflowing gas as the column densities involved increase. If this is a general property of highly obscured nuclei it means that nuclear outflows developed behind the curtain of dust in these sources may be missed by surveys in the far-infrared. On the other hand, in Arp~299, which also contains an obscured nucleus \citep{fal17}, a possible outflow with a projected terminal velocity of approximately $-400$~km\,s$^{-1}$ is seen both in the $1667$~MHz line \citep{baa89} and in the $84$~$\mu$m lines \citep{fal17}. Although this is a possible counterexample, it could also mean that the obscured nucleus in Arp~299 is less extreme than those in Zw~049.057 and Arp~220, or that the outflow in this system is more extended. An alternative reason for the lack of outflow signatures in the far-IR OH lines observed with \emph{Herschel} is that these lines seem to be mostly sensitive to wide-angle outflows \citep{gon17}, possibly due to the limited sensitivity of \emph{Herschel}. Thus, if the outflow is not wide-angle, it is possible that the far-IR observations simply were not able to detect it.

\subsection{Is the outflowing molecular gas entrained by a radio jet?}
\begin{figure}  
  \centering
  \includegraphics[width=8.0cm,trim={0 0 0 1.2cm},clip]{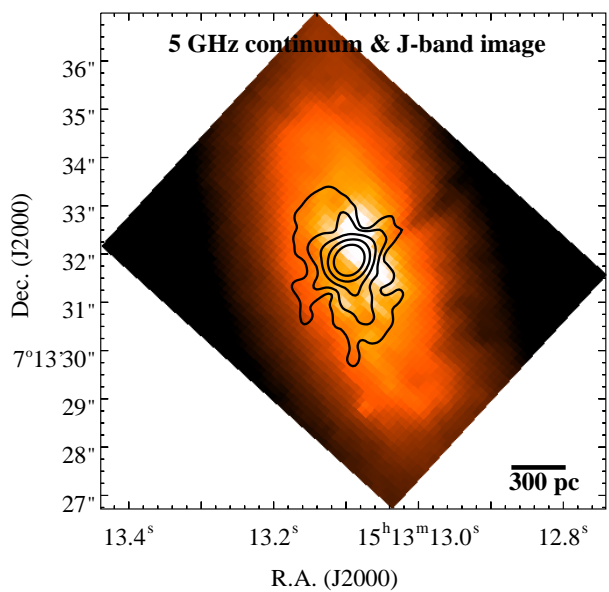}
  \caption{Radio continuum contours overlaid on the HST J-band image. The contours are at $1,4,16,64,256\times15$~$\mu$Jy\,beam$^{-1}$.}
  \label{fig:HST_VLA}
\end{figure}

The radio continuum image of Zw~049.057 (Fig. \ref{fig:VLAcontinuum}) shows one extended component that is aligned with the molecular disk traced by CO (Fig. \ref{fig:co_moms}) as well as one that is inclined with respect to it. This inclined, narrow, component is seen on both sides of the nucleus, with the two parts aligned with each other. In addition, the northwestern part of the feature lines up with a highly collimated, jet-like, dust feature seen in \emph{Hubble} Space Telescope (HST) NICMOS \citep{sco00} and WFC3 (Gallagher, priv. comm.) images. The new HST images will be discussed in detail in a future paper, but we reproduce the J-band image with an overlay of the radio continuum in Fig. \ref{fig:HST_VLA}. Due to the inclination of the galaxy only the dust feature on the side facing us (upper right) is visible, while the radio feature is clearly seen on both sides. In the radio continuum image we can see the possible jet out to ${\sim}350$~pc (${\sim} 1.3\arcsec$) in each direction from the center and the collimated dust feature extends at least twice as far. Assuming that the radio feature is indeed a jet, an estimate of the jet power ($W_{\mathrm{jet}}$) can be made using Eq. (16) of \citet{bir08} relating the jet power to the $1.4$~GHz monochromatic radio power of the total source. According to the measurements of \citet{baa06} the flux density of the total source at $1.4$~GHz is $51$~mJy, yielding a jet power of $1.8\times10^{43}$~erg\,s$^{-1}$, very similar to the estimate made in the same way by \citet{gar14} in NGC~1068. 

\begin{figure}  
  \centering
  \includegraphics[width=8.0cm]{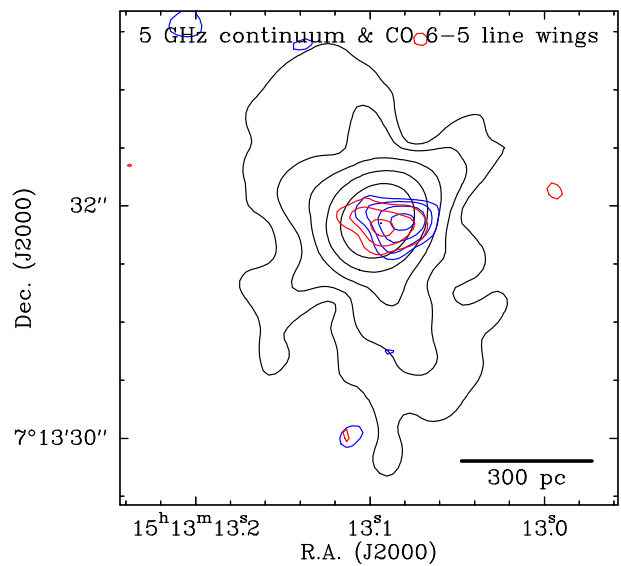}
  \caption{Integrated intensity maps over the line wings ($200-350$~km\,s$^{-1}$ beyond the systemic velocity) in the CO $J=6-5$ emission overlaid on the $5.3$~GHz continuum map (black line).}
  \label{fig:cont_outflow}
\end{figure}

The possible radio jet and the aligned dust feature show that a collimated outflow is emerging from the nucleus of Zw~049.057, but if this is related to the spectral outflow signatures in OH and CO is still unclear at this point. If they are related, the situation might be similar to the one suggested in, for example, M~51 \citep{mat04,mat07}, NGC~3256 \citep{sak14}, and NGC~1266 \citep{ala11} where molecular outflows are attributed to entrainment by a radio jet. This explanation has also been invoked in NGC~1377, where \citet{aal16} suggest that the precessing molecular jet is driven either by a faint radio jet or an accretion disk-wind. On the other hand, a plot of the wings in the CO $J=6-5$ emission overlaid on the radio continuum (see Fig. \ref{fig:cont_outflow}) reveals that the nuclear CO $J=6-5$ outflow (PA ${\sim}105\degr$) does not have the same position angle as the radio feature (PA ${\sim}130\degr$). This can be compared to the minor axis of the disk traced by CO $J=2-1$ which has a position angle of ${\sim}120\degr$. Further studies, with higher spatial resolution and sensitivity, could reveal if this is also the case for the OH and CO $J=2-1$ outflows as well as help estimate the mass outflow rate. This would make it possible to determine if the radio feature is indeed a jet and if it is related to, and able to provide enough energy to drive, the outflow or if another explanation has to be found.

\section{Conclusions}\label{sec:conclusions}
We have used the VLA and the SMA to observe the compact obscured nucleus of Zw~049.057 at radio and millimeter wavelengths. In addition, we have also used data from ALMA and HST. The SMA observations reveal that the line profile of the CO $J=2-1$ emission at the position of the nucleus has wings extending to $\pm300$~km\,s$^{-1}$ from the systemic velocity. The limited spatial resolution (${\sim}1\arcsec=270$~pc) of the SMA observations precludes a direct determination of whether these wings are a signature of outflowing or rapidly rotating gas in the nucleus. However, ALMA observations of CO $J=6-5$ at $0.2\arcsec$ resolution reveal similar line wings which are spatially separated along an axis approximately perpendicular to the overall velocity field of the galaxy. In addition, VLA observations of the $4.7$ and $6.0$~GHz OH lines in absorption at $0.3\arcsec-0.4\arcsec$ resolution show that they all have blue wings extending down to $-300$~km\,s$^{-1}$ from the systemic velocity, supporting the outflow interpretation of the CO line profile. These outflow signatures are not seen in the OH absorption lines previously observed with \emph{Herschel}, suggesting that nuclear outflows in very obscured galaxies might be missed by far-IR observations.

The radio continuum at $5$~GHz is partly resolved and reveals, for the first time, a ${\sim}450$~pc long radio feature that seems to be associated with a highly collimated dust feature seen in an HST J-band image. At this point it is unclear exactly how the radio and the dust features are related to the spectral outflow signature seen in CO and OH, but there are indications that they are oriented in approximately the same direction. Further studies of both the radio feature and the molecular outflow could reveal how they are related to each other and to the nuclear activity, as well as how this activity affects the host galaxy.

By doing a simple comparison of the apparent optical depths of the OH main lines, we find that the excitation conditions in Zw~049.057 likely differ from those in other OH megamasers. A possible explanation for this difference is that Zw~049.057 has entered a transition phase between the OH megamasers and the weaker kilomasers, although a more extensive study would be required to confirm this. 

\begin{acknowledgements}
  The Submillimeter Array is a joint project between the Smithsonian Astrophysical Observatory and the Academia Sinica Institute of Astronomy and Astrophysics and is funded by the Smithsonian Institution and the Academia Sinica.
  This paper makes use of the following ALMA data: ADS/JAO.ALMA\#2013.1.00524.S. ALMA is a partnership of ESO (representing its member states), NSF (USA) and NINS (Japan), together with NRC (Canada), MOST and ASIAA (Taiwan), and KASI (Republic of Korea), in cooperation with the Republic of Chile. The Joint ALMA Observatory is operated by ESO, AUI/NRAO and NAOJ.
  NF and SA thank the Swedish National Space Board for generous grant support (grant numbers 145/11:1B, 285/12 and 145/11:1-3).
  KS is supported by grant MOST 106-2119-M-001-025.
  GCP was supported by a FONDECYT Postdoctoral Fellowship (No. 3150361).
  This research has made use of NASA's Astrophysics Data System (ADS) and of GILDAS software (http://www.iram.fr/IRAMFR/GILDAS).
  \end{acknowledgements}

\bibliographystyle{bibtex/aa} 
\bibliography{ref} 

\end{document}